\def\year{2020}\relax
\documentclass[letterpaper]{article} % DO NOT CHANGE THIS
\usepackage{aaai20}  % DO NOT CHANGE THIS
\usepackage{times}  % DO NOT CHANGE THIS
\usepackage{helvet} % DO NOT CHANGE THIS
\usepackage{courier}  % DO NOT CHANGE THIS
\usepackage[hyphens]{url}  % DO NOT CHANGE THIS
\usepackage{graphicx} % DO NOT CHANGE THIS
\usepackage{graphicx}  % DO NOT CHANGE THIS
\usepackage{xspace}
\usepackage{graphicx}
\usepackage{amsmath}
\usepackage{booktabs}
\usepackage{xcolor}
\usepackage{xspace}
\usepackage{dsfont}
\usepackage{comment}
\usepackage{multirow}
\usepackage{makecell}

\usepackage{cuted}
\usepackage{capt-of}

\newcommand{\projtitle}{How Useful Are the Machine-Generated Interpretations to General Users?\\A Human Evaluation on Guessing the Incorrectly Predicted Labels}
\title{\projtitle}

\newcommand{\citet}[1]{\citeauthor{#1} \shortcite{#1}} 
\newcommand{\citep}{\cite}

\newcommand{\smap}{saliency map\xspace}
\newcommand{\smaps}{saliency maps\xspace}

\newcommand{\vint}{visual interpretation\xspace}
\newcommand{\vints}{visual interpretations\xspace}

\newcommand{\dnn}{deep neural network\xspace}
\newcommand{\dnns}{deep neural networks\xspace}
\newcommand{\wrongly}{incorrectly\xspace}

\author{Hua Shen,~Ting-Hao (Kenneth) Huang\\
College of Information Sciences and Technology\\
The Pennsylvania State University\\
201 Old Main, University Park, PA 16802, USA\\
\{huashen218,~txh710\}@psu.edu
}

\begin{document}
\maketitle

%------------------- Abstract -------------------
\begin{abstract}
Explaining to users why automated systems make certain mistakes is important and challenging.
Researchers have proposed ways to automatically produce interpretations for \dnn models.
However, it is unclear how \textit{useful} these interpretations are in helping users figure out why they are getting an error.
If an interpretation effectively explains to users how the underlying \dnn model works, people who were presented with the interpretation should be better at predicting the model's outputs than those who were not.
This paper presents an investigation on whether or not showing machine-generated \vints helps users understand the \textbf{incorrectly predicted labels} produced by image classifiers.
We showed the images and the correct labels to 150 online crowd workers and asked them to select the incorrectly predicted labels with or without showing them the machine-generated \vints.
The results demonstrated that displaying the visual interpretations did not increase, but rather \textit{decreased}, the average guessing accuracy by roughly 10\%.
\end{abstract}

\begin{figure*}[t]
    \centering
    \includegraphics[width=.95\textwidth]{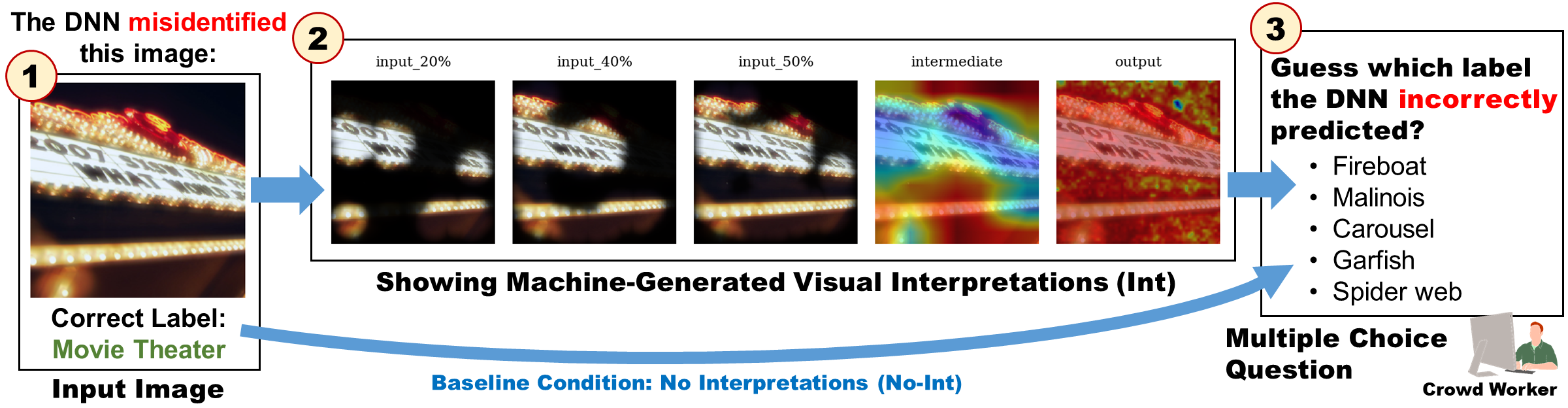}
    \vspace{-.5pc}
    \caption{The workflow of ``Guessing the Incorrectly Predicted Label'' task. Each worker is presented with an image and told that the \dnn incorrectly predicted its label (Step 1). The worker may also be presented with \vints (Step 2). The worker is then asked to guess the incorrectly predicted label (``Carousel'' in this example) from five options, four of them being distractors (Step 3). If an interpretation effectively explains how the underlying \dnn model works to users, the people who were presented with the interpretation should be better at predicting the model's outputs.}
    \label{fig:workflow}
    \vspace{-.7pc}
\end{figure*}

%------------------- Introduction -------------------
\section{Introduction}
\label{sec:introduction}

Explaining to users why automated systems make certain mistakes is important.
As \dnn technologies achieve higher performance, they have been applied to important domains, influencing important decisions in healthcare, transportation, and education.
However, due to the non-linear, complicated structures of neural models, the high performance of \dnns is achieved at the cost of interpretability. 
In response, researchers have proposed ways to explain the inner workings of \dnns by automatically producing interpretations ~\cite{faithfulness:2018:nips,selvaraju:gradcam,lime}.
Such machine-generated interpretations help various stakeholders~\cite{lstmvis:2017:ieee}:
researchers, who develop new deep-learning architectures;
machine-learning engineers, who train and optimize existing networks;
product engineers, who apply general-purpose pre-trained networks to various tasks;
and the general users, who want to understand system outputs~\cite{visual:2020:arxiv,no:2020:chi,selvaraju:gradcam}.
This paper focuses on the \textbf{end users} -- who may not understand the mechanism of the underlying \dnns, but are most influenced by their outputs -- to investigate whether machine-generated interpretations can help users make sense of errors made by algorithms.

We use the image-classification task as our test bed.
Neural image classifiers generate interpretations through two approaches:
designing proxies, which are inherently interpretable ({\em e.g.}, decision tree), to substitute the black-box \dnns~\cite{faithfulness:2018:nips};
or generating post-hoc interpretations outside the \dnn workflow~\cite{selvaraju:gradcam}, which is where our work will focus.
Most post-hoc interpretations are in the form of instance-wise interpretation -- for example, \smaps of input images.
A \smap highlights the most informative region of the image with respect to its classification label, unveiling post-hoc evidence of the neural network prediction.
This line of work was in part motivated by the need of
``end users''~\cite{du:2018:kdd,effects:2019:hcomp}, ``non-expert users''~\cite{lime}, or ``untrained users''~\cite{selvaraju:gradcam}, and the generated interpretations were often evaluated by how much they could boost users' trust of \dnns.
However, it is still unclear how \textbf{useful} these interpretations are in helping users make sense of automated system errors.

The need for interpretability arises due to \textit{Incompleteness} in the problem formalization, making it difficult to make further judgements or optimizations~\cite{doshi:2017:towards}.
When a user observed a few cases where the automated system incorrectly labeled his/her images, it was difficult for the user to decide what to do.
Did the errors occur because the system's accuracy level is low?
If so, should the user switch to another system?
Are the images too complicated for computers, in which case users should not expect reliable image labels?
Did the underlying algorithms have biases that worsened with certain types of images?
We argue that errors \textit{expose} existing incompleteness in the problem formalization, requiring users to seek interpretations.
Namely, an important use case of interpretations is to help users figure out what is going on when they get certain errors.
Researchers have proposed evaluations to assess how much an interpretation reflects the model's behavior (also known as ``fidelity'')~\cite{faithfulness:2018:nips}
or boosts users' trust in automated systems~\cite{selvaraju:gradcam,lime}.
However, it is unclear how \textit{useful} these interpretations are in helping users figure out why they are getting an error.

This paper introduces a method that uses crowd workers from Amazon Mechanical Turk (MTurk) to directly evaluate the usefulness of interpretations in helping users to reason about the errors of \dnns\footnote{The code and interface are available via GitHub:\\\url{https://github.com/huashen218/GuessWrongLabel}}.
Figure~\ref{fig:workflow} overviews the workflow.
In this task, each worker is presented with an image and told that the \dnn incorrectly predicted its label.
The worker may also be presented with a set of interpretations ({\em e.g.}, \smaps) that explain how the \dnn ``perceives'' this image and makes the final prediction.
The worker is then asked to \textbf{guess the incorrectly predicted label} from five options, four of them being distractors.
If an interpretation effectively explains how the underlying \dnn model works to users, the people who were presented with the interpretation should be better at predicting the model's outputs than those who were not.

This paper tried to answer two research questions:
First \textbf{(RQ1)}, do machine-generated \vints help human users better identify predicted labels?
Second \textbf{(RQ2)}, when do (and when do not) the \vints help?

%------------------- Literature -------------------
\section{Related Work}
\label{sec:realted-work}

\paragraph{Interpretation Methods} 
Our work focuses on post-hoc interpretations.
These methods generate \smaps to indicate where the neural networks ``look'' in the images for their predictions' evidence.
Existing methods can be categorized into four lines:
\emph{Backprop-Based}: computes the gradient (or variants) of the neural network output to score the importance of each input pixel, such as SmoothGrad~\citep{smoothgrad};
\emph{Representation-Based}: uses the feature maps at intermediate layer of neural networks to generate \smaps, like GradCAM~\citep{selvaraju:gradcam}; 
\emph{Meta-Model-Based}: trains a meta-model to predict the \smap for any given input in a single feed-forward pass, such as RTS~\citep{Dabkowski:nips:2017}; 
\emph{Perturbation-Based}: finds the \smap by perturbing the input with minimum intervention and observing the change in model prediction, like ExtremalPerturb~\citep{fong:2019:iccv}.

\paragraph{Evaluating Interpretations}
Evaluating the effectiveness of interpretations is critical in practice.
Existing evaluations answer two questions: whether the interpretations genuinely reflect neural network behavior~\cite{Adebayo:nips:2018}, and whether the interpretations are useful for users.
To answer the latter question, a set of metrics are proposed to involve human evaluation.
For instance, trust assessment and user satisfaction is verified in~\citet{no:2020:chi} by surveying general users.
Mental model evaluations designed by~\citet{proxy:2020:iui} and
\citet{visual:2020:arxiv} measure whether general users can understand and predict model outputs.
\citet{can:2019:iui} creates a human-computer cooperative task to measure how much interpretation improves human performance.
However, more study is needed to investigate how general users perceive and predict neural networks' failure cases, which is of vital importance in building trust and correcting model behavior.

\paragraph{Human-AI Collaboration}
Although human computation has traditionally played a data annotation role in deep learning systems, there is increasing interest in incorporating it into diverse stages of human-AI hybrid systems~\citep{effects:2019:hcomp}.
Due to its goal of building human understanding and trust in black-box neural networks, interpretation is inherently a human-centric problem.
Related efforts involve human perception of different types of interpretation representations in visual interfaces~\citep{roy2019explainable}, etc.

%------------------- Method -------------------
\section{Method}
\label{sec:method}

We used a \dnn to label images and employed several interpreters to generate \vints for the images.
We showed each image the \dnn had labeled incorrectly to a group of online crowd workers and asked them to guess which images the \dnn had mistakenly labelled.
Only the workers in the control group were presented with the \vints.
We detail the procedure of the study in this section.

\paragraph{Step 1: Labeling Images}
We trained an image classifier on ImageNet dataset, with its TOP-1 accuracy reaching 78.67\%~\cite{xie2019unsupervised}.
We randomly selected images whose labels were incorrectly identified by the classifier.

\begin{figure}[t]
    \centering
    \includegraphics[width=\columnwidth]{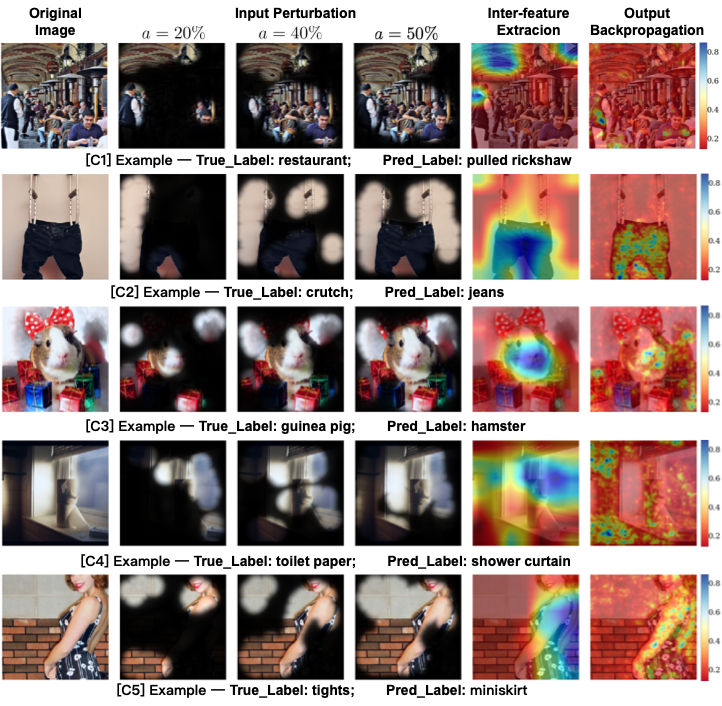}
    \vspace{-2pc}
    \caption{Examples of five types of errors in image classification. The \vints are generated by three existing interpreters (see ``Step 2'' in the Method section.)}
    \label{fig:multi_view}
    \vspace{-1pc}
\end{figure}

\paragraph{Step 2: Generating Instance-Wise Interpretations}
For each image in the misclassified subset, we used three existing interpreters -- {\em i.e.,} input perturbation~\cite{fong:2019:iccv}, intermediate feature extraction~\cite{selvaraju:gradcam}, and output backpropagation~\cite{smoothgrad} -- to explain three aspects of this image.
\textbf{Input perturbation interpretation (column 2-4 in Figure~\ref{fig:multi_view})} observes how the output value changes as input is ``deleted'' in different sub-regions.
We used \emph{ExtremalPerturb}, which aims to find a small pixel subset that, when preserved, are sufficient to keep model output stable.
Moreover, ExtremalPerturb allows researchers to explicitly constrain the percentage of preserved pixels.
We provided three levels of percentage: $a=\{20\%, 40\%, \text{and }50\%\}$. 
\textbf{Inter-Feature extraction interpretation (column 5 in Figure~\ref{fig:multi_view})} looks at intermediate layers of the neural network to indicate the discriminative image regions used by the model for prediction.
We used \emph{GradCAM}, which extracts the gradient information flowing into the last convolutional layers, to explain the importance of each pixel.
\textbf{Output backpropagation interpretation (column 6 in Figure~\ref{fig:multi_view})} leverages backpropagation to track information from the model's output back to its input to generate the \smap.
We used \emph{SmoothGrad}, which samples similar images by adding noise to the original image and using the average of the resulting heatmaps to obtain the final interpretation.
We eventually generated {\em (i)} three \smaps from input perturbation view with 20\%, 40\% and 50\% percentages respectively, {\em (ii)} one \smap from intermediate feature extraction view, and {\em (iii)} one \smap from the output backpropagation view.

\paragraph{Step 3: Having Crowd Workers Guess the Incorrectly Predicted Label}
Next, we recruited crowd workers on MTurk to complete tasks\footnote{Each Human Intelligence Task (HIT) contained one image, and multiple workers were recruited to answer the question. The price of a HIT is \$0.05. Four built-in MTurk qualifications are used: Locale (US Only), HIT Approval Rate ($\geq$98\%), Number of Approved HITs ($\geq$3000), and the Adult Content Qualification.}.
The workers were shown the image and its correct label, and were informed that ``a computer algorithm misidentified this image as something else.''
Only the workers in the control group, as shown in Figure~\ref{fig:workflow}, were presented with the \vints.
On the interface, we explained that the \vints are ``visualizations that try to show how the algorithm \textit{sees} this image,'' and provided comprehensive descriptions for each interpretation.
For example, we explained ``input perturbation interpretation'' with a 20\% mask (column 2 in Figure~\ref{fig:multi_view}) as ``We only allow the algorithm to see 20\% of the image and ask the algorithm to choose which 20\% is the most important region. The black mask blocks the regions the algorithm pays less attention to.''
The workers are then asked to \textbf{guess the \wrongly predicted label} from five options. 
One of the options was the incorrect label predicted by the \dnn model, and the remaining four were randomly selected from the whole label set of ImageNet ({\em i.e.,} 1,000 labels), excluding the correct gold-standard label.

The assumption is that if the \vints effectively explain how the \dnn works, the workers who were presented with the interpretations should distinguish the predicted label better than those who were not.
Humans alone are sufficient to guess the \textit{correct} label, but it requires workers to take the mechanism of \dnns into account to guess the \textit{incorrect} label predicted by \dnns.
MTurk workers are appropriate participants because they represent general users who do not necessarily understand \dnn models nor are trained for reasoning about these models' errors.

\paragraph{Categorizing Error Cases Manually}
To inspect usefulness of interpretation in fine-grained model failure scenarios (RQ2), the authors inspected 1,000 misclassified images and categorized them into five types of errors (Figure~\ref{fig:multi_view}), in part based on the literature~\cite{irm:2019:arxiv}.

\begin{enumerate}
    \item
    \textbf{Local Character Inference (C1):} 
    The model arrives at wrong prediction by looking at only part of the object. For instance, in Figure~\ref{fig:multi_view}(C1), the error might be due to the model partially capturing the restaurant dome, which looks similar to the canopy of a pulled rickshaw.

    \item 
    \textbf{Multiple Objects Selection (C2):} For images with multiple objects, the model makes a prediction by choosing another object rather than the labeled one, as in Figure~\ref{fig:multi_view}(C2).

    \item 
     \textbf{Similar Appearance Inference (C3):} The model misclassifies the object in the image into another class with a similar appearance, as shown in Figure~\ref{fig:multi_view}(C3).

    \item
    \textbf{Correlation Learning (C4):} The model exploits correlational relationships in training data to apply an incorrect label to the image. For example, in Figure~\ref{fig:multi_view}(C4), the model predicts a ``shower curtain'' by identifying the bathroom context, even if no curtain is in the image.
    
    \item 
    \textbf{Incorrect Gold-Standard Labels (C5):} The true label of the images might be incorrect in the ImageNet. Figure~\ref{fig:multi_view}(C5) shows an example.
    
\end{enumerate}

%------------------- Experiment 1 -------------------
\section{Experimental Results}

\paragraph{Experiment 1: Testing Two Conditions in the Same Batch of HITs}
Experiment 1 had two conditions: [Interpretation] ({\em i.e.}, [Int]) and [No-Interpretation] ({\em i.e.}, [No-Int]). 
The only difference is that HITs in the [No-Int] group do not show the interpretations to workers in interfaces.
We evenly divided 200 randomly selected image samples into two groups.
We posted these 200 images in a same batch of HITs at the same time on MTurk, where each HIT recruits nine different workers.
A total of 1,800 submissions (900 submissions in each condition) were contributed by 41 workers in [Int] and 40 workers in [No-Int] conditions respectively.
We did not control the workers' participation, so a worker could participate in both groups.
Thirty-six out of 45 workers participated in both conditions.

Surprisingly, in Experiment 1, \textbf{showing the workers machine-generated \vints \textit{reduced} their average accuracy in guessing the incorrectly predicted labels.}
We calculated the accuracy as the percentage of correctly inferring the classifier's prediction among all 900 submissions in each condition.
The accuracy collected in [Int] was 0.73, while the accuracy in [No-Int] was 0.81.
The difference was statistically significant (unpaired t-test, p$<$0.05, N=100).
Based on the results, the machine-generated interpretation did not help, but instead hurt, the workers' ability to guess the \wrongly predicted labels.
The by-category analysis (Table~\ref{tab:expt1_accuracy}) shows that displaying interpretations significantly lowers human accuracy in cases where the errors were probably caused by similar appearances between items (C3) or by mistakenly learning from the background or scenes of the image (C4).

\begin{table}[t]
\centering
\small
\begin{tabular}{@{}lrrrrrr@{}}
\toprule
 & \textbf{C1} & \textbf{C2} & \textbf{C3} & \textbf{C4} & \textbf{C5} & \textbf{Overall} \\ \midrule
\textbf{Int} & 0.77 & 0.83 & 0.71 & 0.54 & 0.71 & 0.73 \\
\textbf{\#images} & 29 & 23 & 28 & 15 & 5 & 100 \\ \midrule
\textbf{No-Int} & 0.76 & 0.77 & \textbf{**0.87} & \textbf{**0.75} & 0.78 & \textbf{*0.81} \\
\textbf{\#images} & 25 & 10 & 47 & 12 & 6 & 100 \\ \bottomrule
\end{tabular}
\caption{Results of Experiment 1. Showing the workers machine-generated \vints \textit{reduced} their average accuracy in guessing the incorrectly predicted labels. (Unpaired t-test. *: p$<$0.05, **: p$<$0.01.)}
\label{tab:expt1_accuracy}
\vspace{-.8pc}
\end{table}

%------------------- Experiment 2 -------------------
\paragraph{Experiment 2: Testing with Two None Overlapping Sets of Workers}

Experiment 2 was controlled more strictly.
We randomly selected another 200 images (different from those used in Experiment 1), and used the same photo in both [Int] and [No-Int] conditions.
We used custom MTurk qualifications to control the participants: workers who participated in one condition could not accept HITs in the other condition.
We recruited 10 different workers for each image, in which five workers were in the [Int] group and the other five were in the [No-Int] group.
A total of 2,000 submissions (with 1,000 submissions in each condition) were collected, contributed by 42 workers in the [Int] condition and 63 workers in the [No-Int] condition respectively.

In Experiment 2, the machine-generated \vint again \textbf{reduced the average human accuracy in inferring model misclassification} (Table~\ref{tab:expt2_accuracy}.)
The accuracy of [Int] was 0.63, whereas accuracy in [No-Int] condition was 0.73.
The difference was again statistically significant (paired t-test, p$<$0.01, N=200).
On average, humans do not benefit from interpretations when inferring incorrect predictions in image classification tasks.
Similarly to Experiment 1, the by-category analysis showed that displaying interpretations significantly lowers human accuracy in C3 and C4 (Table~\ref{tab:expt2_accuracy}) errors.
We also noticed that the accuracy for C1 and C2 images increased in both experiments when showing \vints, although the differences were not statistically significant.

\begin{table}[t]
\small
\centering
\begin{tabular}{@{}lrrrrrr@{}}
\toprule
 & \textbf{C1} & \textbf{C2} & \textbf{C3} & \textbf{C4} & \textbf{C5} & \textbf{Overall} \\ \midrule
\textbf{Int} & 0.57 & 0.74 & 0.66 & 0.41 & 0.67 & 0.63 \\
\textbf{No-Int} & 0.52 & 0.71 & \textbf{**0.84} & \textbf{*0.59} & 0.77 & \textbf{**0.73}  \\
\textbf{\#images} & 44 & 20 & 112 & 18 & 6 & 200 \\ \bottomrule
\end{tabular}
\caption{Results of Experiment 2. The machine-generated \vint again \textit{reduced} the average human accuracy in inferring model misclassification. (Paired t-test. *: p$<$0.05, **: p$<$0.01.)}
\label{tab:expt2_accuracy}
\vspace{-.8pc}
\end{table}

%------------------- Discussion -------------------
\section{Discussion}
\label{sec:discussion}

Our experiments showed that, in the case of image classification, machine-generated \vints are not necessarily useful in helping users understand \dnn failures.
It could even be harmful, as in the cases where the errors were probably caused by similar appearances between items (C3) or by mistakenly learning from the background or scenes of the images (C4).
System designers should use caution when displaying machine-generated interpretations to users.

\paragraph{Why it did not help?}
More research is required to discover why showing interpretations was ineffective.
Here, we submit several of our hypotheses with the goal of helping future explorations.
First, the interpreters are not good enough to help humans.
The representational power -- including the correctness, sensitivity, etc., of the interpretation model -- might not be sufficient to augment human reasoning about errors.
Although machine-generated interpretations captured some of the \dnn's behaviors, it may not be good enough to help humans.
Second, the format is insufficient. The \smaps may not be the most efficient format to convey information to humans.
For example, when a \smaps model changes an inner parameter, this change might not be obvious enough to be noticeable by humans, but could still affect the final predictions.
Third, the interpreters may work poorly in cases where the image classifier failed.

\paragraph{Limitations}
We are aware that this work has several limitations.
First, the sample size was relatively small.
Given that classifiers incorrectly labelled more than 10,000 images in the ImageNet validation set alone, 200 images are relatively small portion of the data.
Second, we only tested three particular types of interpretations, and also presented the interpretations together on the same page.
This experimental setup
introduces the possibility of missing out on the ``best'' interpretations, or different interpretations might affect each other and reduce their effectiveness.
Third, we recruited MTurk workers with certain qualifications to simulate general users.
It is difficult to eliminate data noise stemmed from workers' misunderstanding or incognizance of images or options.
Finally, we only tested \vints for image classifiers.
It requires more research to study if similar effects could be generalized to other tasks.

%------------------- Conclusion -------------------
\section{Conclusion}
\label{sec:conclusion}
The goal of this study was to evaluate the usefulness of machine-generated \vints for general users' reasoning about model errors.
To this end, we utilized the ``guess \wrongly predicted labels'' task to examine the usefulness of \vints.
Our two sets of control experiments, with 3,800 submissions contributed by 150 online crowd workers, suggest that showing the interpretations does not increase, but rather {\em decreases}, the average accuracy of human guesses by roughly 10\%.

%------------------- Acknowledgements -------------------
\section{Acknowledgements}
We thank Ting Wang for his support.
We also thank the workers on MTurk who participated in our studies.

%------------------- Bibliography -------------------
\bibliographystyle{aaai.bst}
\bibliography{bibliography.bib}

\newpage

\end{document}